# The non-significance factor is a simple posterior estimate of the minimum necessary sample size.

I Novikov, I. Tessler, A. Yakirevich.

Abstract.

A researcher is interested in what sample size is needed to get the required significance of the same test, assuming exactly the same situation that was in the study with the non-significant result. We propose a simple solution to the problem.

Problem.

It is well known that many studies find a non-significant but not null effect. The test used may be, for example, a t-test, or a likelihood ratio test, or a Wald test for some parameters in very complex regression models. A fundamental practical question is "what sample size is needed to get the significance of the same test, assuming exactly the same situation that was in the collected data." "Situation" involves the same model, the same null hypothesis, the same test, same sampling design, and the same joint distribution of all parameters in the data set. This condition rules out any effect of random sampling. Hence, there is no room for power. The situation differs from the well-known problem determining the sample size using data from a pilot study [e.g. 1]. The most natural interpretation of "exactly the same situation" is that each observation will be repeated the same number of times, and the same model and test will be used. Thus, the question can be reformulated as: "How many times should I observe each observation to get a given significance?" This number of times is what we call the "non-significance factor" (NF).

Solution.

Find the smallest frequency weight W such that the same test on an original sample of N observations with that weight W would result in a significance less than or equal to the required significance p.

Most programs only accept integer weights. In this situation, a more accurate non-integer value of $W_{int}$ can be obtained by linear interpolation of the largest W0, which still results in a 'non-significant' p-value p0 > p, and the smallest integer W1, which results in a "significant" p-value of p1<p:

$W_{int}$=(W0*(p-p1) + W1*(p0-p)) /(p0-p1).

One can apply other interpolations using more values of the weight W with the corresponding p-values and various interpolation approaches. However, we believe that the more complex models with their assumptions do not justify the illusion of a more accurate estimate.

It is important to note that NF depends not only on the observed p-value, but also on the situation as a whole, including initial sample size, joint distribution of parameters, model and test. We provide an example demonstrating this property of NF.

Example.

We compared NF for identical p-values for the Cox multiple regression likelihood ratio test and the Wald intercept test in linear regression without covariates. The data for the Cox regression is the first 30 observations from the STATA example "https://www.stata-press.com/data/r16/stan3". An example for linear regression was built to produce the same p-value in a sample of 30 observations from a standard normal distribution. The full STATA output is in the Appendix.

The observed p-values were 0.6433 for the Cox regression and 0.643 for the Wald test. As a result, the NFs were:

Wald test: W0=17, N0=510, p0=0.05, $N_{int}$=510.

Cox: W0=4, N0=120, p0=0.0826; W1=5, N1=150, p1=0.0392; $W_{int}$=4.751, $N_{int}$=142.53

We see that NFs are quite different and lead to different sample sizes.

Discussion.

The non-significance factor NF is an a posteriori measure. It is different from "posterior power", which is another metric for a similar purpose based solely on the observed p-value. The posterior power has been heavily criticized for several reasons [eg 2]. We believe that the main disadvantages of posterior power is that it depends on the observed significance only and provides the researcher with little information. On the contrary, NF depends not only on the observed p-value, but also on the situation as a whole. This is an accurate answer to the natural question "how much sample size is needed to obtain a given significance for the model and test, if everything exactly matches the collected data."

The non-significance factor from the usual sample size estimates required for a given significance and power based on a pilot study. The advantage of NF is that it can be used for any joint distribution of all study parameters and any complex model and test. In contrast, all standard calculations of the required sample size use strict assumptions about the data and test. The exception is the bootstrap approach, which uses the same "same situation" idea. Bbootstrap does not reproduce the situation exactly. It considers random samples from the available data and gives the estimate of the sample size necessary to provide a given power. If the current data set was too small, it can be non-representative and then bootstrap can lead to biased estimates of the sample size.

Conclusion.

The non-significance factor NF is a simple a posteriori estimate of the sample size required to obtain a given significance "if the situation were exactly the same". However, it should be emphasized that NF cannot be used directly when planning a new study. It does not take into account the effect of randomness and therefore should be considered as a lower estimate of the sample size required to provide a given significance of the same test on the sample from the same situation..

References.

Appendix. STATA output.

```
.   use https://www.stata-press.com/data/r16/stan3, clear

(Heart transplant data)

.   drop if _n>30

(142 observations deleted)

.   keep t1 died id age posttran surg year
```

```
. list

     +------------------------------------------------------+
     | id   year   age   died   surgery   posttran     t1 |
     |------------------------------------------------------|
  1. |  1     67    30      1         0          0     50 |
  2. |  2     68    51      1         0          0      6 |
  3. |  3     68    54      0         0          0      1 |
  4. |  3     68    54      1         0          1     16 |
  5. |  4     68    40      0         0          0     36 |
     |------------------------------------------------------|
  6. |  4     68    40      1         0          1     39 |
  7. |  5     68    20      1         0          0     18 |
  8. |  6     68    54      1         0          0      3 |
  9. |  7     68    50      0         0          0     51 |
 10. |  7     68    50      1         0          1    675 |
     |------------------------------------------------------|
 11. |  8     68    45      1         0          0     40 |
 12. |  9     68    47      1         0          0     85 |
 13. | 10     68    42      0         0          0     12 |
 14. | 10     68    42      1         0          1     58 |
 15. | 11     68    47      0         0          0     26 |
     |------------------------------------------------------|
 16. | 11     68    47      1         0          1    153 |
 17. | 12     68    53      1         0          0      8 |
 18. | 13     68    54      0         0          0     17 |
 19. | 13     68    54      1         0          1     81 |
```

```
 20. | 14      68      53        0           0            0      37 |
     |---------------------------------------------------------------|
 21. | 14      68      53        1           0            1    1386 |
 22. | 15      68      53        1           0            0       1 |
 23. | 18      68      56        0           0            0      20 |
 24. | 18      68      56        1           0            1      43 |
 25. | 17      68      20        1           0            0      36 |
     |---------------------------------------------------------------|
 26. | 16      68      49        0           0            0      28 |
 27. | 16      68      49        1           0            1     308 |
 28. | 19      68      59        1           0            0      37 |
 29. | 20      69      55        0           0            0       1 |
 30. | 20      69      55        1           0            1      28 |
     +---------------------------------------------------------------+
```

. gen ww=1

. stset t1 [fweight=ww], failure(died) id(id)

```
              id:  id
    failure event:  died != 0 & died < .
obs. time interval:  (t1[_n-1], t1]
 exit on or before:  failure
           weight:  [fweight=ww]
```

---------------------------------------------------------------------

```
        30  total observations
         0  exclusions
```

```
      --------------------------------------------------------------
            30   physical observations remaining, equal to
            30   weighted observations, representing
            20   subjects
            20   failures in single-failure-per-subject data
         3,071   total analysis time at risk and under observation
                                 at risk from         t =          0
                           earliest observed entry t =          0
                                last observed exit t =      1,386

. foreach vv of numlist 1 4 5 {
.   display " *************************weight= `vv'***************"
.   replace ww=`vv'
.   stset t1 [fweight=ww], failure(died) id(id)
.   stcox age posttran surg year
. }
 *************************weight= 1*******************
(0 real changes made)

             id:  id
  failure event:  died != 0 & died < .
obs. time interval:  (t1[_n-1], t1]
 exit on or before:  failure
         weight:  [fweight=ww]

--------------------------------------------------------------
            30   total observations
```

```
        0   exclusions
-----------------------------------------------------------------
       30   physical observations remaining, equal to
       30   weighted observations, representing
       20   subjects
       20   failures in single-failure-per-subject data
    3,071   total analysis time at risk and under observation
                                at risk from t =         0
                       earliest observed entry t =         0
                            last observed exit t =     1,386

      failure _d:  died
 analysis time _t: t1
             id:  id
         weight:  [fweight=ww]

note: surgery omitted because of collinearity
Iteration 0:   log likelihood = -42.335616
Iteration 1:   log likelihood =  -41.51509
Iteration 2:   log likelihood = -41.499967
Iteration 3:   log likelihood = -41.499959
Refining estimates:
Iteration 0:   log likelihood = -41.499959

Cox regression -- no ties

No. of subjects =            20       Number of obs    =           30
```

```
No. of failures =                20       Time at risk    =           3071
                                          LR chi2(3)      =           1.67
Log likelihood  =    -41.499959           Prob > chi2     =         0.6433

------------------------------------------------------------------------------
       _t |  Haz. Ratio   Std. Err.      z    P>|z|     [95% Conf. Interval]
----------+-------------------------------------------------------------------
      age |   .9763896    .0269655    -0.87   0.387     .9249432    1.030697
  posttran |     .61161    .3984063    -0.75   0.450      .170607    2.192564
   surgery |          1  (omitted)
      year |    2.90105    3.001411     1.03   0.303      .381862    22.03961
------------------------------------------------------------------------------

************************weight= 4*******************
(30 real changes made)

              id:  id
   failure event:  died != 0 & died < .
obs. time interval:  (t1[_n-1], t1]
 exit on or before:  failure
          weight:  [fweight=ww]

------------------------------------------------------------------------------
      30  total observations
       0  exclusions
------------------------------------------------------------------------------
      30  physical observations remaining, equal to
     120  weighted observations, representing
```

```
        80  subjects
        80  failures in single-failure-per-subject data
    12,284  total analysis time at risk and under observation
                                    at risk from t =          0
                              earliest observed entry t =     0
                               last observed exit t =     1,386

             failure _d:  died
       analysis time _t:  t1
                    id:  id
                weight:  [fweight=ww]

note: surgery omitted because of collinearity

Iteration 0:   log likelihood = -280.24601

Iteration 1:   log likelihood = -276.96391

Iteration 2:   log likelihood = -276.90342

Iteration 3:   log likelihood = -276.90339

Refining estimates:

Iteration 0:   log likelihood = -276.90339

Cox regression -- no ties

No. of subjects =              80      Number of obs     =         120

No. of failures =              80      Time at risk      =       12284

                                       LR chi2(3)        =        6.69

Log likelihood  =    -276.90339        Prob > chi2       =      0.0826
```

```
------------------------------------------------------------------
      _t |  Haz. Ratio   Std. Err.      z    P>|z|   [95% Conf. Interval]
---------+--------------------------------------------------------
     age |   .9763896    .0134827    -1.73   0.084    .9503183    1.003176
 posttran |    .61161    .1992031    -1.51   0.131    .3230247    1.158013
  surgery |         1   (omitted)
     year |   2.90105    1.500706     2.06   0.040    1.052521    7.996125
------------------------------------------------------------------

************************weight= 5********************

(30 real changes made)

               id:  id
    failure event:  died != 0 & died < .
obs. time interval:  (t1[_n-1], t1]
  exit on or before:  failure
           weight:  [fweight=ww]

------------------------------------------------------------------
        30   total observations
         0   exclusions
------------------------------------------------------------------
        30   physical observations remaining, equal to
       150   weighted observations, representing
       100   subjects
       100   failures in single-failure-per-subject data
    15,355   total analysis time at risk and under observation
                                          at risk from t =         0
```

```
                            earliest observed entry t =           0
                               last observed exit t =       1,386

        failure _d:  died
   analysis time _t:  t1
                id:  id
            weight:  [fweight=ww]

note: surgery omitted because of collinearity

Iteration 0:   log likelihood = -372.62187
Iteration 1:   log likelihood = -368.51924
Iteration 2:   log likelihood = -368.44362
Iteration 3:   log likelihood = -368.44359
Refining estimates:
Iteration 0:   log likelihood = -368.44359

Cox regression -- no ties

No. of subjects =           100            Number of obs    =         150
No. of failures =           100
Time at risk    =         15355
                                           LR chi2(3)       =        8.36
Log likelihood  =    -368.44359            Prob > chi2      =      0.0392

------------------------------------------------------------------------
     _t | Haz. Ratio   Std. Err.      z    P>|z|   [95% Conf. Interval]
--------+---------------------------------------------------------------
```

```
        age |   .9763896   .0120593    -1.93   0.053    .9530375    1.000314
   posttran |     .61161   .1781727    -1.69   0.091     .345545    1.082541
    surgery |          1  (omitted)
       year |    2.90105   1.342272     2.30   0.021    1.171432    7.184446
------------------------------------------------------------------------------

.
. gen x=.
(30 missing values generated)

. set seed 137248

. forvalues nn=1(1) 30 {
  2.   quietly     replace x=rnormal() in `nn'
  3. }

. gen y=x-0.0455

. gen wtt=1

.  foreach vv of numlist 1 16 17 18 {
.     display " **************** ttest *****weight= `vv'************"
.     replace wtt=`vv'
.     regress y [fw=wtt]
. }
 *************** ttest *****weight= 1 *******************
(0 real changes made)
```

```
      Source |       SS           df       MS      Number of obs   =        30
-------------+------------------------------       F(0, 29)        =      0.00
       Model |           0         0       .       Prob > F        =         .
    Residual |  34.1462048        29  1.17745534   R-squared       =    0.0000
-------------+------------------------------       Adj R-squared   =    0.0000
       Total |  34.1462048        29  1.17745534   Root MSE        =    1.0851

------------------------------------------------------------------------------
           y |      Coef.   Std. Err.      t    P>|t|     [95% Conf. Interval]
-------------+----------------------------------------------------------------
       _cons |   .0929164   .1981124     0.47   0.643    -.3122689    .4981017
------------------------------------------------------------------------------

*************** ttest *****weight= 16 ********************
(30 real changes made)

      Source |       SS           df       MS      Number of obs   =       480
-------------+------------------------------       F(0, 479)       =      0.00
       Model |           0         0       .       Prob > F        =         .
    Residual |  546.339276       479  1.14058304   R-squared       =    0.0000
-------------+------------------------------       Adj R-squared   =    0.0000
       Total |  546.339276       479  1.14058304   Root MSE        =     1.068

------------------------------------------------------------------------------
           y |      Coef.   Std. Err.      t    P>|t|     [95% Conf. Interval]
-------------+----------------------------------------------------------------
       _cons |   .0929164   .0487464     1.91   0.057    -.0028669    .1886997
```

*************** ttest *****weight= 17 ********************

(30 real changes made)

```
      Source |       SS           df       MS      Number of obs   =        510
-------------+------------------------------      F(0, 509)       =       0.00
       Model |          0            0        .   Prob > F        =          .
    Residual |  580.485481          509 1.14044299  R-squared      =     0.0000
-------------+------------------------------      Adj R-squared   =     0.0000
       Total |  580.485481          509 1.14044299  Root MSE       =     1.0679

------------------------------------------------------------------------------
           y |      Coef.   Std. Err.      t    P>|t|     [95% Conf. Interval]
-------------+----------------------------------------------------------------
       _cons |   .0929164   .0472881     1.96   0.050     .0000125    .1858202
------------------------------------------------------------------------------
```

*************** ttest *****weight= 18 ********************

(30 real changes made)

```
      Source |       SS           df       MS      Number of obs   =        540
-------------+------------------------------      F(0, 539)       =       0.00
       Model |          0            0        .   Prob > F        =          .
    Residual |  614.631686          539 1.14031853  R-squared      =     0.0000
-------------+------------------------------      Adj R-squared   =     0.0000
       Total |  614.631686          539 1.14031853  Root MSE       =     1.0679

------------------------------------------------------------------------------
```

```
      y |      Coef.   Std. Err.     t    P>|t|    [95% Conf. Interval]
---------+----------------------------------------------------------------
   _cons |   .0929164   .0459532    2.02   0.044    .002647    .1831858
--------------------------------------------------------------------------
```

.
end of do-file

.